\documentstyle[12pt,epsf]{article}

\begin{document}

\begin{titlepage}

\setcounter{footnote}{1}
\setcounter{page}{1}

\vbox{}
\flushright{MIT-CTP-3007, astro-ph/0101507}
\vskip 1.5 truein
\begin{center}
{\Large \bf Eternal Inflation\footnote{Talk given at {\it Cosmic
Questions,} April 14-16, 1999, National Museum of Natural
History, Washington, D.C., organized by the Program of Dialogue
on Science, Ethics, and Religion of the American Association for
the Advancement of Science.  To appear in the proceedings (The
New York Academy of Sciences Press).}\\}
\bigskip \bigskip
{\bf Alan H. Guth\footnote{This work is supported in part by
funds provided by the U.S. Department of Energy (D.O.E.) under
cooperative research agreement \#DF-FC02-94ER40818.}\\}
\medskip
{\small \it Center for Theoretical Physics\\} {\small \it
Laboratory for Nuclear Science and Department of Physics\\}
{\small \it Massachusetts Institute of Technology, Cambridge,
Massachusetts\ \ 02139\ \ \ U.S.A.\\}
\bigskip
{\tt guth@ctp.mit.edu\\}
\bigskip \bigskip
\end{center}

\begin{abstract}
The basic workings of inflationary models are summarized, along
with the arguments that strongly suggest that our universe is the
product of inflation.  It is argued that essentially all
inflationary models lead to (future-)eternal inflation, which
implies that an infinite number of pocket universes are produced. 
Although the other pocket universes are unobservable, their
existence nonetheless has consequences for the way that we
evaluate theories and extract consequences from them. The
question of whether the universe had a beginning is discussed but
not definitively answered.  It appears likely, however, that
eternally inflating universes do require a beginning.
\end{abstract}

\end{titlepage}

\setcounter{footnote}{0}

\section{Introduction}
\setcounter{equation}{0}

     The question of whether or not the universe had a beginning
is, of course, by no means an easy question.  When you ask a
scientist a question that is not easy, he never gives just one
answer, but instead gives a succession of answers.  In this case,
I would like to offer two levels of answers.

     At the first level, I would argue that the answer to the
question is yes, the universe had a beginning in the event that
is usually referred to as the big bang.

     I think that at least 99.9 percent of the people working in
scientific cosmology today believe that the universe evolved from
a hot dense state, exactly as Sandra Faber spoke about earlier. 
This theory is strongly supported by the direct observation of
the expansion of the universe via the redshift of the light from
distant galaxies, by the measurement of the abundances of the
light chemical elements, and by the now very precise measurements
of both the spectrum and the very small nonuniformities of the
cosmic microwave background radiation.  Thus, most scientists
(including me) believe that the universe as we know it began in a
``big bang'' some 11 to 16 billion years ago \cite{general}. 

     However, as Sandra has already emphasized, there is another
level to the question of beginning.  When cosmologists say that
they are persuaded that the big bang theory is valid, they are
using a rather precisely defined and restricted interpretation of
the term ``big bang.'' As it is used by scientists the term
refers only to the expansion of the universe from an initially
hot dense state.  But it says nothing about whether the universe
really began there, or whether there was something else that
preceded what we call the big bang.

     So, beyond the standard big bang, there is now a very
significant body of research concerning the possibility of cosmic
inflation \cite{Guth1,Linde1,Albrecht-Steinhardt1,generalinf}. 
Today I want to talk about inflation, and in particular I want to
talk about a very likely ramification called eternal inflation. 
As you will see, the theory of inflation does not give a clear
answer to the question of whether the universe had a beginning,
but it does provide at least a context for discussing this
question.

     To begin, I would like to highlight the distinction between
the questions that the standard big bang theory answers and the
questions that inflation is intended to answer.

     The standard big bang theory is, of course, a very
significant scientific theory.  It describes how the early
universe expanded and cooled from an initially very hot dense
state.  It describes how the light chemical elements that we
observe today were synthesized during the first 200 seconds or so
of this expansion period.  And finally, although work in this
area is still in progress, it seems to describe very well how the
matter in the universe eventually congealed to form the stars,
galaxies, and clusters that we observe in the universe today.

     There is, however, a key issue that the standard big bang
theory does not discuss at all: it does not tell us what banged,
why it banged, or what happened before it banged.  Despite its
name, the big bang theory does not describe the bang at all.  It
is really only the theory of the {\bf aftermath} of a bang.

     So, in particular, the standard big bang theory does not
address the question of what caused the expansion; rather, the
expansion of the universe is incorporated into the equations of
the theory as an assumption about the initial state---the state
of the universe when the theory begins its description.

     Similarly, the standard big bang theory says nothing about
where the matter in the universe came from.  In the standard big
bang theory all the matter that we see here, now, was already
there, then.  The matter was just very compressed, and in a form
that is somewhat different from its present state.  The theory
describes how the matter evolved from one form to another as the
universe evolved, but the theory does not address the question of
how the matter originated.

     While inflation does not go so far as to actually describe
the ultimate origin of the universe, it does attempt to provide a
theory of the bang: a theory of what it was that set the universe
into expansion, and at the same time supplied essentially all of
the matter that we observe in the universe today.

\section{How Does Inflation Work?}
\setcounter{equation}{0}

     I will begin by giving a quick rundown of how inflation
works.  Some of these issues were already discussed by Sandra
Faber, who I thought gave an excellent description.  For
completeness, however, I will start my explanation at the
beginning, but I will try to go more quickly when discussing
points that Sandra has already explained.

     The key idea---the underlying physics---that makes inflation
possible is the fact that most modern particle theories predict
that there should exist a state of matter that turns gravity on
its head, creating a gravitational repulsion.  This state can
only be reached at energies well beyond those that we can probe
experimentally, but the theoretical arguments for the existence
of the state are rather persuasive.  It is not merely the
prediction of some specific theory, but it is the generic
prediction for a wide class of plausible theories.  Thus, gravity
does not always have to be attractive.\footnote{The possibility
of repulsive gravity arises because, according to Einstein's
theory of general relativity, gravitational fields are produced
not just by energy or mass densities, but also by pressures.  The
direction of the field caused by pressure is what you would
probably guess: a positive pressure---the kind that we normally
see---produces an attractive gravitational field.  But the
peculiar state of matter that I'm talking about produces a
negative pressure, which you might also call a suction.  It is in
fact a very large negative pressure, resulting in a repulsive
gravitational field which is stronger than the attractive field
produced by the mass density of the matter.  The result is a net
gravitational repulsion, which is the driving force behind
inflation.}

     The gravitational repulsion caused by this peculiar kind of
material is the secret behind inflation. Inflation is the
proposition that the early universe contained at least a small
patch that was filled with this peculiar repulsive-gravity
material.\footnote{The name for this peculiar gravitationally
repulsive state is not well-established.  Sandra Faber referred
to it as a vacuum with a finite energy density, and sometimes as
a false vacuum.  In my own technical articles I call it a false
vacuum, although I have to explain that for most inflationary
models this usage stretches the meaning for which the phrase had
previously been used in particle physics.  In this article I will
refer to it as a repulsive-gravity material.  It may seem strange
to see the words ``vacuum'' and ``material'' used to describe the
same thing, but keep in mind that this stuff is {\bf strange.}
The word ``vacuum'' is used to emphasize that it is different
from ordinary matter, while I am calling it a material to
emphasize that it is different from an ordinary vacuum!} There
are a variety of theories about how this might have happened,
based on ideas ranging from chaotic initial conditions to the
creation of the universe as a quantum tunneling event.  Despite
the ambiguity of this aspect of the theory, there are two things
to keep in mind.  First, the probability of finding a region
filled with this repulsive-gravity material need not be large.  I
will come back to this point later, and argue that it is only
necessary that the probability is nonzero.  Second, the resulting
predictions do not depend on how the initial patch was formed. 
Once the patch exists, inflation takes over and produces a
universe that ends up inevitably looking very much like the one
that we live in.

     The initial patch can be incredibly small.  It need be only
about one-billionth the size of a single proton.  Once the patch
exists it starts to rapidly expand because of its internal
gravitational repulsion.  The expansion is exponential, which
means it is characterized by a doubling time, which for a typical
inflationary theory might be in the neighborhood of $10^{-37}$
seconds.  So every $10^{-37}$ seconds the diameter of the patch
doubles, and then it doubles again and again during each
$10^{-37}$-second interval.  The success of the description
requires about a hundred of these doublings, but there could have
been many more.  In the course of this expansion, the patch went
from being a tiny speck to a size at least as large as a marble.

     So the patch of repulsive-gravity material expanded by a
huge factor.  Whenever a normal material expands its density goes
down, but this material behaves completely differently.  As it
expands, the density remains constant.  That means that the total
amount of mass contained in the region increased during inflation
by a colossal factor.

     The increase in mass probably seems strange at first,
because it sounds like a gross violation of the principle of
energy conservation.  Mass and energy are equivalent, so we are
claiming that the energy of the matter within the patch increased
by a colossal factor.  The reason this is possible is that the
conservation of energy has a sort of a loophole, which physicists
have known at least since the 1930s \cite{tolman}, but haven't
talked about very much.  Energy is always conserved; there are no
loopholes to that basic statement.  However, we normally think of
energies as always being positive.  If that were true, then the
large amount of energy that we see in the universe could not
possibly have gotten here unless the universe started with a lot
of energy.  However, this is the loophole: energies are not
always positive.  In particular, the energy of a gravitational
field is negative.  This statement, that the energy of a
gravitational field is negative, is true both in the context of
the Newtonian theory of gravity and also in the more
sophisticated context of general relativity.

     So, during inflation, total energy is conserved.  As more
and more {\bf positive} energy (or mass) appears as the patch
expands at constant density, more and more {\bf negative} energy
is simultaneously appearing in the gravitational field that fills
the region.  The total energy is constant, and it remains
incredibly small because the negative contribution of gravity
cancels the enormous positive energy of the matter.  The total
energy, in fact, could very plausibly be zero.  It is quite
possible that there is a perfect cancellation between the
negative energy of gravity and the positive energy of everything
else.

     For the theory to be successful, there has to be a mechanism
to end the period of inflation---the period of accelerated
expansion---because the universe is not undergoing inflation
today.\footnote{Actually there is strong evidence that the
expansion of the universe is accelerating in the present era, and
the mechanism for this acceleration is believed to be very
similar to that of inflation.  This acceleration, however, is
much slower than the acceleration that inflationary models
propose for the early universe, so in any case the rapid
acceleration of the early universe must have come to an end.}
Inflation ends because the repulsive-gravity material is
fundamentally unstable.  So it doesn't survive forever, but
instead decays like a radioactive substance.  Like traditional
forms of radioactive decay, it decays exponentially, which means
that the decay is characterized by a half-life.  During any
period of one half-life, on average half of the repulsive-gravity
material will ``decay'' into normal attractive-gravity material.

     In the process of decaying, the repulsive-gravity material
releases the energy that has been locked up within itself.  That
energy evolves to become a hot soup of ordinary particles. 
Initially the decay produces a relatively small number of
high-energy particles, but these particles start to scatter off
of each other.  Eventually the energy becomes what we call {\it
thermalized,} which means that it produces an equilibrium gas of
hot particles---a hot primordial soup---which is exactly the
initial condition that had always been assumed in the context of
the standard big bang theory.

     Thus, inflation is an add-on to the standard big bang
theory.  Inflation supplies the beginning to which the standard
big bang theory then becomes the continuation.

\section{Evidence for Inflation}
\setcounter{equation}{0}

     So far I have tried to describe how inflation works, but now
I would like to explain the reasons why many
scientists---including certainly myself---believe that inflation
really is the way that our observed universe began.  There are
six reasons that I will discuss, starting with some very general
ideas and then moving to more specific ones.

     The first reason is the obvious statement that the universe
contains a tremendous amount of mass.  It contains about
$10^{90}$ particles within the visible region of the universe.  I
believe that most non-scientists are somewhat puzzled to hear
anyone make a fuss over this fact, since they think, ``Of course
the universe is big---it's the whole universe!'' However, to a
theoretical cosmologist who is hoping to build a theory to
explain the origin of the universe, this number seems like it
could be an important clue.  Any successful theory of the origin
of the universe must somehow lead to the result that it contains
at least $10^{90}$ particles.  The fundamental theory on which
the calculation is based, however, presumably does not contain
any numbers nearly so large.  Calculations can of course lead to
factors of 2 or $\pi$, but it would take very many factors of 2
or $\pi$ to reach $10^{90}$.  Inflation, however, leads to
exponential expansion, and that seems to be the easiest way to
start with only small numbers and finish with a very large one. 
With inflation the problem of explaining why there are $10^{90}$
or more particles is reduced to explaining why there were 100 or
more doubling times of inflation.\footnote{Since the volume is
proportional to the cube of the diameter, during 100 doublings
the volume increases by a factor of $\left(2^{100}\right)^3 =
2^{300} \approx 2 \times 10^{90}$.} The number 100 is modest
enough so that it can presumably arise from parameters of the
underlying particle physics and/or geometric factors, so
inflation seems like just the right kind of theory to explain a
very large universe.

     The second reason is the Hubble expansion itself---the fact
that the universe is observed to be in a state of uniform
expansion.  An ordinary explosion, like TNT or an atomic bomb,
does not lead to expansion that is nearly uniform enough to match
the expansion pattern of the universe.  But the gravitational
repulsion of inflationary models produces exactly the uniform
expansion that was first observed by Edwin Hubble in the 1920s
and 30s.

     Third, inflation is the only theory that we know of that can
explain the homogeneity and isotropy of the universe---that is,
the uniformity of the universe.  This uniformity is observed most
clearly by looking at the cosmic microwave background radiation,
which we view as the afterglow of the heat of the big bang.  The
intensity of this radiation is described by an effective
temperature, and it is observed to have the same temperature in
every direction to an accuracy of about one part in a hundred
thousand, after we correct for our own motion through the cosmic
background radiation.  In other words, this radiation is
incredibly smooth.  As an analogy we can imagine a marble that
has been ground so smoothly that its radius is uniform to one
part in a hundred thousand.  The marble would then be round to an
accuracy of about a quarter of the wavelength of visible light,
about as precise as the best optical lenses that can be
manufactured with present-day technology. 

     In the standard big bang theory there is no explanation
whatever for this uniformity.  In fact, one can even show that
within the context of the standard big bang theory, no
explanation for this uniformity is possible.  To see this, we
need to understand a little about how this cosmic background
radiation originated.  During the first approximately 300,000
years of the history of the universe, the universe was hot enough
so that the matter was in the form of a plasma---that is, the
electrons were separated from the atoms.  Such a plasma is very
opaque to photons, which are constantly scattered by their
interactions with the free electrons.  Although the photons move
of course at the speed of light, they change directions so
rapidly that they essentially go nowhere.  During the first
300,000 years of the history of the universe, the photons were
essentially pinned to the matter.

     But after 300,000 years, according to calculations, the
universe cooled enough so that the plasma neutralized.  The free
electrons combined with the atomic nuclei to form a neutral gas
of hydrogen and some helium, which is very, very transparent to
photons.  From then on these photons have traveled in straight
lines.  So, just as I see an image of you when I observe the
photons coming from your face, when we look at the cosmic
background radiation today we are seeing an image of the universe
at 300,000 years after the big bang.  Thus, the uniformity of
this radiation implies that the temperature must have been
uniform throughout this whole region by 300,000 years after the
big bang.

     To think about whether the temperature of the observed
universe could have equilibrated by this early time, we could try
to imagine fancifully that the universe was populated by little
purple creatures, whose sole purpose in life was to make the
temperature as uniform as possible.  We could imagine that each
purple creature was equipped with a little furnace, a little
refrigerator, and a cell phone so that they could communicate
with each other.  The communication, however, turns out to be an
insurmountable problem.  A simple calculation shows that in order
for them to achieve a uniform temperature by 300,000 years, they
would need to be able to communicate at about a hundred times the
speed of light.  But nothing known to physics allows
communication faster than light, so even with dedicated purple
creatures we could not explain the uniformity of the cosmic
background radiation.

     So in the standard version of the big bang theory, before
inflation is introduced, one simply has to hypothesize that the
universe started out uniform.  The initial uniformity would then
be preserved, since the laws of physics are by assumption the
same everywhere.  This approach allows one to accommodate the
uniformity of the universe, but it is not an explanation.

     Inflation gets around this problem in a very simple way. In
inflationary theories the universe evolves from a very tiny
initial patch.  While this patch was very small, there was plenty
of time for it to become uniform by the same mechanism by which a
slice of pizza sitting on the table cools to room temperature:
things tend to come to a uniform temperature.  Once this
uniformity is established on the scale of the very tiny patch,
inflation can take over and magnify the patch to become large
enough to encompass everything that we observe.  Thus, inflation
provides a very natural explanation for the uniformity of the
universe. 

     Reason number four is known as the flatness problem.  It is
concerned with the closeness of the mass density of our universe
to what cosmologists call the {\it critical density.} The
critical density is best defined as that density which would
cause the universe to be spatially flat.  To understand what this
means, one must understand that, according to general relativity,
the geometry of space is determined by the matter that it
contains.  If the mass density of the universe is very high, the
space will curve back on itself to form a closed universe, the
three-dimensional analogue of the two-dimensional surface of a
sphere.  The sum of the angles in a triangle would exceed
180$^\circ$, and parallel lines would meet if they are extended. 
If the mass density is very low, the space would curve in the
opposite way, forming an open universe.  The sum of the angles in
a triangle would then be less than 180$^\circ$, and parallel
lines would diverge if they were extended.  But with just the
right mass density (for a given expansion rate) the spatial
geometry will be exactly Euclidean, just like what we all learned
in high school---180$^\circ$ in every triangle, and parallel
lines remain parallel no matter how far they are extended.  This
borderline case is the critical density.\footnote{To clear up a
possible source of confusion, I mention that the critical density
has often been described in the semi-popular literature as the
borderline between eternal expansion and eventual collapse.  If
Einstein's cosmological constant is zero, as most of us thought a
few years ago, then this definition is equivalent to the one
given above.  Recent evidence, however, suggests that the
cosmological constant may be nonzero, in which case the two
definitions are not equivalent.  In that case, the one given in
the text agrees with the definition used in the technical
literature, and is also the definition that is relevant to the
current discussion.}

     Cosmologists use the Greek letter $\Omega$ (Omega) to denote
the ratio of the average mass density of the universe to the
critical density:
  $$\Omega \equiv {\hbox{actual mass density} \over
     \hbox{critical mass density}} \ .$$
Today there is growing evidence that $\Omega$ is equal to one to
within about 10\% \cite{deBernardis}, but the issue is not
completely settled.  For purposes of this discussion, therefore,
I will begin with the uncontroversial statement that $\Omega$
lies somewhere in the interval between 0.1 and 2. 

     You would probably not expect that much could be concluded
from such an noncommittal starting point, but in fact it tells us
a lot.  When one looks at the equations describing the evolution
of the universe, it turns out that $\Omega = 1$ is an unstable
equilibrium point, like a pencil balanced on its tip.  If the
pencil is started exactly vertical and stationary, the laws of
Newtonian mechanics imply that it will remain vertical forever. 
But if it is not absolutely vertical, it will rapidly start to
fall in whichever direction it is leaning.  Similarly, if
$\Omega$ began exactly equal to 1, it would remain 1 forever. 
But any deviation from 1 will grow rapidly as the universe
evolves.  Thus, for $\Omega$ to be anywhere in the ballpark of 1
today, it must have started extraordinarily close to 1.  For
example, if we extrapolate backwards to 1 second after the big
bang, $\Omega$ must have been 1 to an accuracy of 15 decimal
places \cite{dicke}.  While 1 second may sound like an
extraordinarily early time at which to be discussing an 11- to
16-billion-year-old universe, cosmologists have pretty much
confidence in the extrapolation.  The nucleosynthesis processes,
which are successfully tested by present measurements of the
abundances of the light chemical elements, where already
beginning at 1 second after the big bang.

     In fact, particle theorists like myself are attempting to
push the history of the universe back to what we call the Planck
time, about $10^{-43}$ seconds, which is the era when quantum
gravity effects are believed to have been important.  Since our
understanding of quantum gravity is still very primitive, we
generally make no attempt to discuss the history of the universe
at earlier times.  But if we attempt to extrapolate the history
of the universe back to the Planck time without invoking
inflation, we find that $\Omega$ at the Planck time must have
been equal to 1 to an accuracy of 58 decimal places.

     Without inflation, there is no explanation for the initial
value of $\Omega$.  The big bang theory is equally
self-consistent for any initial value of $\Omega$, so one has no
a priori reason to prefer one value over another.  But if the
theory is to agree with observation, one must posit an initial
value of $\Omega$ that is extraordinarily close to 1.

     With inflation, on the other hand, during the brief
inflationary era the evolution of $\Omega$ behaves completely
differently.  Instead of being driven away from 1, during
inflation $\Omega$ is driven very strongly towards 1.  So with
inflation you could assume that $\Omega$ started out as 1, 2, 10,
$10^3$, or $10^{-6}$.  It doesn't matter.  As long as there was
enough inflation, then $\Omega$ would have been driven to 1 to
the extraordinary accuracy that is needed.

     The fifth reason for believing the inflationary description
is the absence of magnetic monopoles.  Grand unified particle
theories, which unify all the known particle interactions with
the exception of gravity, predict that there should be stable
particles that have a net magnetic charge.  That is, these
particles would have a net north pole or a net south pole, which
is very different from an ordinary bar magnet which always has
both a north pole and a south pole.  These magnetic monopoles
are, according to our theories, extraordinarily heavy particles,
weighing about $10^{16}$ times as much as a proton.  In the
traditional big bang theory, without inflation, they would have
been copiously produced in the early universe.  If one assumes a
conventional cosmology with typical grand unified theories, one
concludes that the mass density of magnetic monopoles would
dominate all other contributions by an absurdly large factor of
about $10^{12}$ \cite{preskill}.  Observationally, however, we
don't see any sign of these monopoles.

     Cosmic inflation provides a simple explanation for what
happened to the monopoles: in inflationary models, they can
easily be diluted to a negligible density.  As long as inflation
happens during or after the era of monopole production, the
density of monopoles is reduced effectively to zero by the
enormous expansion associated with inflation.

\begin{figure}[ht]
\centerline{\epsfbox{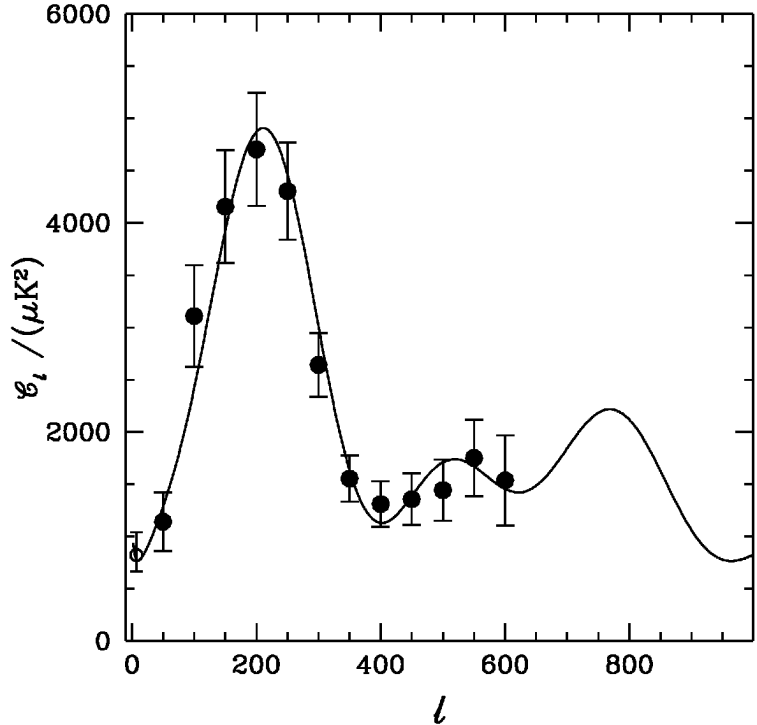}}
\caption{Spectrum of the cosmic background radiation
anisotropies, as measured by the {\sc Boomerang} experiment.  The
intensity of fluctuations is shown as a function of the angular
size parameter $\ell$, where the angular size of a fluctuation is
roughly $180^\circ/\ell$.  The black line is a theoretical curve
corresponding to a standard inflationary model with $\Omega = 1$. 
The mass density in the model is composed of 5\% baryons, 25\%
cold dark matter, and 70\% cosmological constant.  The data and
theoretical curve were taken from Ref.~\protect\cite{boomparam}.}
\label{boomgraph}
\end{figure}

     The sixth and final reason that I would like to discuss for
believing that the universe underwent inflation is the prediction
that the theory makes for the detailed structure of the cosmic
background radiation.  That is, inflation makes very definite
predictions not only for the uniformity that we see around us,
but it also predicts that there should be small deviations from
that uniformity due to quantum uncertainties.  The magnitude of
these deviations depends on the details of the underlying
particle theory, so inflation will not be able to predict the
magnitude until we really understand the particle physics of very
high energies.  However, the shape of the spectrum of these
nonuniformities---i.e., the way that the intensity varies with
wavelength---depends only slightly on the details of the particle
physics.  For typical particle theories, inflationary models
predict something very close to what is called the
Harrison-Zel'dovich, or scale-invariant spectrum.  These
nonuniformities are viewed as the seeds for the formation of
structure in the universe, but they can also be seen directly in
the nonuniformities of the cosmic background radiation, at the
level of about one part in 100,000. Fig.~\ref{boomgraph} shows a
graph of the recent data from the {\sc Boomerang} experiment,
plotted against a theoretical curve derived for an inflationary
model \cite{boomparam}.

\section{Eternal Inflation: Mechanisms}
\setcounter{equation}{0}

     Having discussed the mechanisms and the motivation for
inflation itself, I now wish to move on the main issue that I
want to stress in this article---eternal inflation, the questions
that it can answer, and the questions that it raises.

Before going on, I should clarify that the different topics that
I am discussing have various levels of certainty. The standard
big bang theory, as far as cosmologists are concerned, appears to
be essentially certain to all but a few of us.  Inflation seems
to be by far the most plausible way that the big bang could have
started, but it is not so well established as the big bang
itself.  I should also admit that inflation is vague.  It is not
really a theory, but a class of theories, so there is a
significant amount of flexibility in describing its predictions. 
Eternal inflation, which I am about to describe, seems to me to
be an almost unavoidable consequence of inflation.  This point,
however, is somewhat controversial.  In particular, I believe
that the following speaker, Neil Turok, will argue either that
eternal inflation does not happen, or that it is in any case not
relevant to understanding the properties of the observable
universe.  I, however, will argue that eternal inflation does
happen, and is relevant.

     By eternal inflation, I mean simply that once inflation
starts, it never ends
\cite{vilenkin-eternal,steinhardt-nuffield,linde-eternal}.  The
term ``future-eternal'' would be more precise, because I am not
claiming that it is eternal into the past---I will discuss that
issue at the end of the talk.

     The mechanism that leads to eternal inflation is rather
straightforward to understand.  Recall that we expect inflation
to end because the repulsive-gravity material is unstable, so it
decays like a radioactive substance.  As with familiar
radioactive materials, the decay of the repulsive-gravity
material is generally exponential: during any period of one
half-life, on average half of it will decay.  This case is
nonetheless very different from familiar radioactive decays,
however, because the repulsive-gravity material is also expanding
exponentially.  That's what inflation is all about.  Furthermore,
it turns out that in essentially all models, the expansion is
much faster than the decay.  The doubling-time for the inflation
is much shorter than the half-life of the decay.  Thus, if one
waits for one half-life of the decay, half of the material would
on average convert to ordinary matter.  But meanwhile the part
that remains would have undergone many doublings, so it would be
much larger than the region was at the start.  Even though the
material is decaying, the volume of the repulsive-gravity
material would actually grow with time, rather than decrease. 
The volume of the repulsive-gravity material would continue to
grow, without limit and without end.  Meanwhile pieces of the
repulsive-gravity material decay, producing a never-ending
succession of what I call {\it pocket universes.}

\begin{figure}[ht]
\centerline{\epsfbox{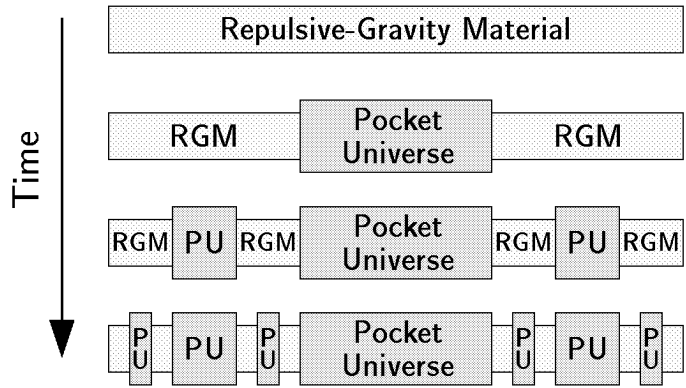}}
\caption{A schematic diagram to illustrate the fractal structure
of the universe created by eternal inflation.  The four
horizontal bars represent a patch of the universe at four evenly
spaced successive times.  The expansion of the universe is not
shown, but each horizontal bar is actually a factor of three
larger than the preceding bar, so each region of
repulsive-gravity material is actually the same size as the
others.  During the time interval between bars, 1/3 of each
region of repulsive-gravity material decays to form a pocket
universe.  The process repeats ad infinitum, producing an
infinite number of pocket universes.}
\label{eternalline}
\end{figure}

     In Fig.~2 I show a schematic illustration of how this works. 
The top row shows a region of repulsive-gravity material, shown
very schematically as a horizontal bar.  After a certain length
of time, a little less than a half-life, the situation looks like
the second bar, in which about a third of the region has decayed. 
The energy released by that decay produces a pocket universe. 
The pocket universe will inflate to become huge, so to its
residents the pocket universe would look like a complete
universe.  But I will call it a pocket universe because there is
not just one, but an infinite number of them. 

     On the second bar, in addition to the pocket universe, we
have two regions of repulsive-gravity material.  On the diagram I
have not tried to show the expansion, because if I did I would
quickly run out of room on the page.  So you are expected to
remember on your own that each bar is actually bigger than the
previous bar, but is drawn on a different scale so that it looks
like it is the same size.  To discuss a definite example, let us
assume that each bar represents three times the volume of the
previous bar.  In that case, each region of repulsive-gravity
material on the second bar is just as big as the entire bar on
the top line.

     The process can then repeat.  If we wait the same length of
time again, the situation will be as illustrated on the third bar
of the diagram, which represents a region that is 3 times larger
than the second bar, and 9 times larger than the top bar. For
each region of repulsive-gravity material on the second bar,
about a third of the region decays and becomes a pocket universe,
leaving regions of repulsive-gravity material in between.  Those
regions of repulsive-gravity material are again just as big as
the one we started with on the top bar. The process goes on
literally forever, producing pocket universes and regions of
repulsive-gravity material between them, ad infinitum.  The
universe on the very large scale acquires a fractal structure. 

     The illustration of Fig.~2 is of course oversimplified in a
number of ways: it is one-dimensional instead of
three-dimensional, and the decays are shown as if they were very
systematic, while in fact they are random.  But the qualitative
nature of the evolution is nonetheless accurate: eternal
inflation really leads to a fractal structure of the universe,
and once inflation begins, an infinite number of pocket universes
are produced.

\section{Eternal Inflation: Implications}
\setcounter{equation}{0}

     The pocket universes other than our own are believed to be
completely unobservable, so one can question whether it makes any
scientific sense to talk about them.  I would argue that it is
valid science, because we are pursuing the consequences of a
theory for which we already have other evidence.  Of course the
theory of inflation has to rest on the evidence that we can
observe, but once we are persuaded by these observations, then I
think that we should also believe the other implications, even if
they involve statements that cannot be directly confirmed.

     If one accepts the existence of the other pocket universes,
then one can still question whether they have any relevance to
the pursuit of science.  I will argue that, even though these
other universes are unobservable, their existence nonetheless has
consequences for the way that we evaluate theories and extract
consequences from them.

     One question for which eternal inflation has relevance is
the question of the ultimate beginning of the universe---what can
be learn about it, and how can we learn it? In the following
talk, Neil Turok will describe his work with Stephen Hawking
\cite{hawking-turok} and others on the origin of the universe as
a quantum event.  He will argue that hypotheses about the form of
the initial wave function lead to statistical consequences for
our universe that can in principle be directly tested.  However,
if eternal inflation is a valid description of the universe (as I
think it is), then I would expect that all such hypotheses about
the ultimate beginning of the universe would become totally
divorced from any observable consequences.  Since our own pocket
universe would be equally likely to lie anywhere on the infinite
tree of universes produced by eternal inflation, we would expect
to find ourselves arbitrarily far from the beginning.  The
infinite inflating network would presumably approach some kind of
a steady state, losing all memory of how it started, so the
statistical predictions for our universe would be determined by
the properties of this steady state configuration, independent of
hypotheses about the ultimate beginning.  In my opinion theories
of the ultimate origin would remain intellectually interesting,
and with an improved understanding of the fundamental laws of
physics, such theories might even eventually become compelling. 
But I expect that any detailed consequences of such a theory
would be completely washed out by the eternal evolution of the
universe.  Thus, there would be no way of relating the properties
of the ultimate origin to anything that we might observe in
today's universe.

Although I believe that the inflating network would approach a
steady state, I should admit that attempts to pursue this idea
quantitatively have run into several technical problems.  First,
the evolution of eternally inflating universes leads to physics
that we do not understand.  In particular, quantum fluctuations
tend to drive the repulsive-gravity material to higher and higher
energy densities, where the poorly understood effects of quantum
gravity become more and more important \cite{LLM,GBLinde}. 
Second, even if we impose enough assumptions so that the
evolution of the eternally inflating universe can be described,
we still do not know how to define probabilities on the infinite
set of pocket universes that is produced \cite{center-world}. 
The problem is akin to asking what fraction of the integers are
odd \cite{odd-integers}.  Most people would presumably say that
the answer is 1/2, since the integers alternate between odd and
even.  However, the ambiguity of the answer can be seen if one
imagines other orderings for the integers.  One could, if one
wished, order the integers as 1,3,\ \ 2,\ \ 5,7,\ \ 4,\ \ 9,11,\
\ 6\ , \dots, always writing two odd integers followed by one
even integer.  This list includes each integer exactly once, but
from this list one would conclude that 2/3 of the integers are
odd.  Thus, the answer seems to depend on the ordering.  For
eternally inflating universes, however, there is no natural
ordering to the regions of spacetime that comprise the entire
universe.  There are well-founded proposals for defining
probabilities \cite{vilenkin-proposal}, but at least in my
opinion there is no definitive and compelling argument. 

     A second implication of eternal inflation is that the
probability for inflation to start---the question of how likely
it is for an initial speck of repulsive-gravity material to
form---becomes essentially irrelevant.  Inflation only needs to
begin once, in all of eternity.  As long as the probability is
nonzero, it does not seem relevant, and perhaps it is not even
meaningful, to ask if the probability is large or small.  If it
is possible, then it will eventually happen, and when it does it
produces literally an infinite number of universes.  Unless one
has in mind some competing process, which could also produce an
infinite number of universes (or at least an infinite space-time
volume), then the probability for inflation to start has no
significance.

     The third and final implication of eternal inflation that I
would like to discuss pertains to the comparison of theories.  I
would argue that once one accepts eternal inflation as a logical
possibility, then there is no contest in comparing an eternally
inflating version of inflation with any theory that is not
eternal.

     Consider the analogy of going into the woods and finding
some rare species of rabbit that has never before been seen.  You
could either assume that the rabbit was created by a unique
cosmic event involving the improbable collision of a huge number
of molecules, or you could assume that the rabbit was the result
of the normal process of rabbit reproduction, even though there
are no visible candidates for the rabbit's parents.  I think we
would all consider the latter possibility to be far more
plausible.  Once we become convinced that universes can eternally
reproduce, then the situation becomes very similar, and the same
logic should apply.  It seems far more plausible that our
universe was the result of universe reproduction than that it was
created by a unique cosmic event.

\section{Did the Universe Have a Beginning?}
\setcounter{equation}{0}

     Finally, I would like to discuss the central topic of this
session, the question of whether or not the universe had a
beginning.

     The name {\it eternal inflation}, as I pointed out earlier,
could be phrased more accurately as {\it future-eternal
inflation.} Everything that has been said so far implies only
that inflation, once started, continues indefinitely into the
future.  It is more difficult to determine what can be said about
the distant past.

     For the explicit constructions of eternally inflating
models, the answer is clear.  Such models start with a state in
which there are no pocket universes at all, just pure
repulsive-gravity material filling space.  So there is definitely
a beginning to the models that we know how to construct.

     In 1993 Borde and Vilenkin \cite{borde-vilenkin} proved a
theorem which showed under fairly plausible assumptions that
every eternally inflating model would have to start with an {\it
initial singularity,} and hence must have a beginning. In 1997,
however, they \cite{borde-vilenkin2} noted that one of their
assumed conditions, although valid at the classical level, was
violated by quantum fluctuations that could be significant in
eternally inflating models.  They concluded that their earlier
proof would not apply to such cases, so the door was open for the
construction of models without a beginning.  They noted, however,
that no such models had been found.

     At the present time, I think it is fair to say that it is an
open question whether or not eternally inflating universes can
avoid having a beginning.  In my own opinion, it looks like
eternally inflating models necessarily have a beginning.  I
believe this for two reasons.  The first is the fact that, as
hard as physicists have worked to try to construct an
alternative, so far all the models that we construct have a
beginning; they are eternal into the future, but not into the
past.  The second reason is that the technical assumption
questioned in the 1997 Borde-Vilenkin paper does not seem
important enough to me to change the conclusion, even though it
does undercut the proof.  Specifically, we could imagine
approximating the laws of physics in a way that would make them
consistent with the assumptions of the earlier Borde-Vilenkin
paper, and eternally inflating models would still exist. 
Although those modifications would be unrealistic, they would not
drastically change the behavior of eternally inflating models, so
it seems unlikely that they would change the answer to the
question of whether these models require a beginning.

     So, as is often the case when one attempts to discuss
scientifically a deep question, the answer is inconclusive. It
looks to me that {\bf probably} the universe had a beginning, but
I would not want to place a large bet on the issue.

\section*{Acknowledgments}
This work is supported in part by funds provided by the U.S.
Department of Energy (D.O.E.) under cooperative research
agreement \#DF-FC02-94ER40818.

\newcommand{\jf}{\it}
\newcommand{\jt}{\/}
\newcommand{\VPY}[3]{{\bf #1}, #2 (#3)}
\newcommand{\ispace}{\thinspace}

\let\U=\.
\def\.{.\nobreak\ispace\ignorespaces}

\newcommand{\IJMODPHYS}[3]{{\jf Int. J. Mod. Phys.\jt} \VPY{#1}{#2}{#3}}
\newcommand{\JETP}[3]{{\jf JETP Lett.\jt} \VPY{#1}{#2}{#3}}
\newcommand{\MPL}[3]{{\jf Mod. Phys. Lett.\jt} \VPY{#1}{#2}{#3}}
\newcommand{\NC}[3]{{\jf Nuovo Cim.\jt} \VPY{#1}{#2}{#3}}
\newcommand{\Nature}[3]{{\jf Nature} \VPY{#1}{#2}{#3}}
\newcommand{\NP}[3]{{\jf Nucl. Phys.\jt} \VPY{#1}{#2}{#3}}
\newcommand{\PHYREP}[3]{{\jf Phys. Rept.\jt} \VPY{#1}{#2}{#3}}
\newcommand{\PL}[3]{{\jf Phys. Lett.\jt} \VPY{#1}{#2}{#3}}
\newcommand{\PR}[3]{{\jf Phys. Rev.\jt} \VPY{#1}{#2}{#3}}
\newcommand{\PRD}[3]{{\jf Phys. Rev. D\jt} \VPY{#1}{#2}{#3}}
\newcommand{\PRL}[3]{{\jf Phys. Rev. Lett.\jt} \VPY{#1}{#2}{#3}}
\newcommand{\PTRSLA}[3]{{\jf Phil. Trans. R. Soc. Lond.\jt\ A}
\VPY{#1}{#2}{#3}}
\newcommand{\ZhETF}[3]{{\jf Zh. Eksp. Teor. Fiz.\jt} \VPY{#1}{#2}{#3}}

\end{document}